\documentclass[pra,twocolumn,amsmath,amssymb]{revtex4-1} %

\usepackage{epsfig,amsmath}
\usepackage{subfigure}
\usepackage{graphicx}% Include figure files
\usepackage{dcolumn}% Align table columns on decimal point
\usepackage{stmaryrd}
\usepackage{mathrsfs}
\usepackage{pifont}
\usepackage{amsthm}
\usepackage{amssymb}
\usepackage{bm}
\usepackage{latexsym}
\usepackage{hyperref}
\usepackage{color}

\newcommand{\la}{\langle}
\newcommand{\ra}{\rangle}

\newcommand{\si}{\sigma}

\newcommand{\om}{\omega}
\newcommand{\Om}{\Omega}

\newcommand{\non}{\nonumber}
\newcommand{\pa}{\partial}

\def\pra#1{{ Phys.\ Rev. A\/} {\bf#1}}

\def\pre#1{{ Phys.\ Rev. E\/} {\bf#1}}
\def\prl#1{{ Phys.\ Rev.\ Lett.} {\bf#1}}

\def\annph#1{{ Ann.\ Phys.} {\bf #1}}
\def\pla#1{{ Phys.\ Lett. A\/} {\bf#1}}
\def\rmp#1{{ Rev. \ Mod. \ Phys.} {\bf#1}}

\begin{document}

%\title{Noise Induced Adiabaticity  }
\title{One Component Dynamical Equation and Noise Induced Adiabaticity}

\author{Jun Jing$^{1,2}$\footnote{Email address: jingjun@shu.edu.cn}, Lian-Ao Wu$^{2,3}$\footnote{Email address: lianao_wu@ehu.es}, Ting Yu$^{4}$, J. Q. You$^{5}$, %J. G. Muga $^{7}$, 
Zhao-Ming Wang$^{6}$, and Lluc Garcia$^{2}$}

\affiliation{$^{1}$Institute of Theoretical Physics and Department of Physics, Shanghai University, Shanghai 200444, China \\ $^{2}$Department of Theoretical Physics and History of Science, The Basque Country University (EHU/UPV), PO Box 644, 48080 Bilbao, Spain \\ $^{3}$Ikerbasque, Basque Foundation for Science, 48011 Bilbao\\ $^{4}$Center for Controlled Quantum Systems and Department of Physics and Engineering Physics, Stevens Institute of Technology, Hoboken, New Jersey 07030, USA \\ $^{5}$Beijing Computational Science Research Center, Beijing 100084, China 
%\\ $^{7}$ Department of Physical Chemistr, The Basque Country University (EHU/UPV), PO Box 644, 48080 Bilbao, Spain 
\\ $^{6}$Department of Physics, Ocean University of China, Qingdao 266100, China}

\date{\today}

%%%%%%%%%%%%%%%%%%%%
\begin{abstract}
The adiabatic theorem addresses the dynamics of a target instantaneous eigenstate of a time-dependent Hamiltonian.
% Multiple instantaneous eigenstates are involved in the dynamical process when the adiabatic conditions are not satisfied. 
We use a Feshbach P-Q partitioning technique to derive a closed one-component integro-differential equation. The resultant equation properly traces the footprint of the target eigenstate. The physical significance of the derived dynamical equation is illustrated by both general analysis and concrete examples.  Surprisingly, we find an anomalous phenomenon showing that a dephasing white noise can enhance and even induce adiabaticity.  This new phenomenon may naturally occur in many physical systems. We also show that white noises can also shorten the total duration of dynamic processes such as adiabatic quantum computing.
\end{abstract}


\maketitle

%\section{Introduction}
{\em Introduction.}---The adiabatic principle is a fundamental concept in quantum mechanics addressing 
quantum evolution governed by a slowly-varying Hamiltonian \cite{Born}. It states that an initial eigenstate of the 
time-dependent Hamiltonian remains in the same instantaneous eigenstate at a later time.  Because of its simplicity, 
this principle has a variety of applications in quantum physics. The recent development in quantum information processing has reinforced the importance of the adiabatic principle by  its wide-spread applications such as quantum adiabatic algorithm \cite{Farhi}, fault-tolerance against quantum errors \cite{Childs}, 
and  universal adiabatic and holonomic quantum computation \cite{Aharonov,Zanardi,Carollo} based on the Berry's phase \cite{Berry,Zee,Barry}. 
 The adiabatic principle also has important applications in quantum dynamics control such as adiabatic passage \cite{Oreg,Bergmann,Kral,Hu}, 
 adiabatic gate teleportation \cite{Bacon} and many other protocols (e.g., see \cite{Nielsen,Lidar,Goswami,Wang,Muga,Wu}). In all those applications,  the choice of 
 initial and the target instantaneous eigenstates is varied depending on the interest of the issues under consideration.

Adiabaticity for a closed system is an idealization. In reality, all experimentally accessible systems are open because of inevitable interactions between systems and their surrounding environments \cite{Gardiner,Breuer}.  While extensive work has been done in using  the closed system adiabaticity combined with an external control mechanism \cite{Oreg,Bergmann,Kral,Hu,Muga},  adiabaticity has been theoretically extended into the context of open quantum systems \cite{Lidar} where the environmental noises often modify or even ruin a designed adiabatic passage. We therefore ask ourselves:  How does a noise affect quantum adiabaticity?  Can adiabaticity be protected or even created by an environmental noise?
% Our primary motivation in this Letter is to study the impact of a noise on the adiabaticity.

Our new discovery is rather surprising; %we find a white noise can have a profound impact on the adiabaticity.  
contrary to our intuition, we show that an external white noise can be used to enhance adiabaticity.  Moreover, we show that a white noise can even induce adiabaticity from a non-adiabatic regime. To put our new discovery into perspective, we present the adiabatic theorem in a slightly different way by noting that the theorem essentially addresses the dynamics of {\em one} target instantaneous eigenstate or one component of the eigenvectors. By using the %powerful 
Feshbach P-Q partitioning technique \cite{Wu09}, we can derive a simple one-component integro-differential equation governing the target instantaneous eigenstate. The derived one-compoment dynamical equation can signal the onset of the adiabaticity if the integrand appearing in the integro-differential equation has a fast-varying factor, whether natural or engineered, such that  the integral in the equation will be small (or zero). Therefore this term's  contribution to the dynamics may be ignored \cite{JW}. 
This is the condition of adiabaticity.  As to be shown below, a particular type of white noise can effectively induce the desired fast-varying factor so that adiabaticity can be established even when the system is originally in a non-adiabatic regime.

{\em One-component dynamical equation derived from Feshbach PQ-partitioning technique.}---We consider a time-dependent Hamiltonian $H(t)$.\nobreak\hspace{.16667em plus .13333em}The instantaneous eigenequation of $H(t)$ is,
\begin{equation}
H(t)|E_n(t)\ra=E_n(t)|E_n(t)\ra,
\end{equation}
where $E_n$'s and $|E_n\ra$'s are instantaneous eigenvalues and non-degenerate eigenvectors, respectively. A quantum state at $t$ can be expressed as $|\psi(t)\ra=\sum_n\psi_n(t)e^{i\theta_n(t)}|E_n(t)\ra$, where the dynamical phase is given by  $\theta_n(t)\equiv-\int_0^tE_n(s)ds$. Substituting it into the Schr\"odinger equation, we obtain the following differential equation, %for the coefficient $\psi_0$,
\begin{equation}\label{cm}
\pa_t \psi_m=-\la E_m|\dot{E}_m\ra \psi_m-\sum_{n\neq m}\la E_m|\dot{E}_n\ra e^{i(\theta_n-\theta_m)}\psi_n,
\end{equation}
where $\la E_m|\dot{E}_n\ra=\frac{\la E_m|\dot{H}|E_n\ra}{E_n-E_m}$  ($n\neq m$). Without loss of generality, the target component can be denoted as $\psi_0$. Throughout the paper, $\psi_0(t)$ corresponds to the target eigenstate of $H(t)$, which is one of eigenvectors $|E_n(t)\ra$. The adiabatic theorem is valid if $|\la E_m|\dot{E}_n\ra|\ll  |E_n-E_m|$. The coefficient of the adiabatic wave function is then  $\psi_0(t)=e^{i\beta_0(t)}$,  where $\beta_0(t)\equiv i\int_0^t\la E_0(s)|\dot{E}_0(s)\ra ds$ is the geometric phase. Physically, the adiabatic theorem asserts that an initial eigenstate  $|E_0(0)\ra$ remains  the target instantaneous eigenstate $|E_0(t)\ra$ at a later time. Equation~(\ref{cm}) is also 
 the Schr\"odinger equation with the effective ``rotating representation" Hamiltonian $H_{mn}=-i\la E_m|\dot{E}_n\ra e^{i(\theta_n-\theta_m)}$ which contains multiple variables $\psi_m$'s \cite{Wuzy}.

As shown above that the adiabatic theorem determines the dynamics of {\em one} target component $\psi_0$, therefore, an exact dynamical equation for the target component $\psi_0(t)$  is highly desirable.  By using the P-Q partitioning technique, $\psi_0$ is shown to satisfy the following integro-differential equation,

\begin{equation}\label{ME1}
\pa_t\psi_0(t)=-\la E_0|\dot{E}_0\ra \,\psi_0(t)-\int_0^tds\,g(t,s)\psi_0(s),
\end{equation}
where $g(t,s)=R(t)G(t,s)W(s)$ is an effective propagator and $G(t,s)=\mathcal{T}_{\leftarrow}\{\exp[-i\int_s^tD(s')ds']\}$ is a time-ordered evolution operator. 
Here the vector $R\equiv[R_1, R_2, \cdots]$ with $R_m=-i\la E_0|\dot{E}_m\ra e^{i(\theta_m-\theta_0)}$, and $W=R^\dag$. The matrix $D\equiv\sum_{mn}D_{mn}|m\ra\la n|$, where $D_{mn}=-i\la E_m|\dot{E}_n\ra e^{i(\theta_n-\theta_m)}$ ($m,n\geq1$). The first term on the right-hand side of (\ref{ME1}) is the same as that in equation (\ref{cm}), 
which corresponds to the Berry's phase that may be switched off in a rotating frame. $|\psi_0(t)|^2$, the probability of finding the eigenstate $|E_0(t)\ra$ at time $t$, is determined by the history of product of the propagator $g(t,s)$ and $\psi_0(s)$.

With the exact dynamical equation (\ref{ME1}), the crucial adiabatic condition can be cast into the following compact form,
\begin{equation}\label{ME2}
\int_0^tds\,g(t,s)\psi_0(s) =0.
\end{equation}
The condition is satisfied when the factor $g(t,s)$ is a rapid oscillating function \cite{Barry}. Physically, it can be easily seen that the average of the product of the fast-varying $g(t,s)$ and the slow-varying $\psi_0(s)$ gives rise to zero. Clearly, the well-known adiabatic  condition corresponds to the first-order approximation of this exact result.

{\em Qubit example and shortcut to adiabaticity.}---Consider a two-level system (TLS) or qubit model in the rotating representation,
\begin{equation}
H(t)=\left(\begin{array}{cc}
      -i\la E_0|\dot{E}_0\ra & -i\la E_0|\dot{E}_1\ra e^{i\int_0^tE(s)ds} \\
      -i\la E_1|\dot{E}_0\ra e^{-i\int_0^tE(s)ds} &
      -i\la E_1|\dot{E}_1\ra
    \end{array}\right)
\end{equation}
where $E\equiv E_0-E_1$. When the TLS is initially in the eigenstate $|E_0\ra$, the propagator $g(t,s)$ is given by,
\begin{equation}\label{Pt}
g(t,s)=-\la E_0(t)|\dot{E}_1(t)\ra\la E_1(s)|\dot{E}_0(s)\ra e^{\int_s^t(iE-\la E_1|\dot{E}_1\ra)ds'}.
\end{equation}

\begin{figure}[htbp]
\centering
\includegraphics[width=2.8in]{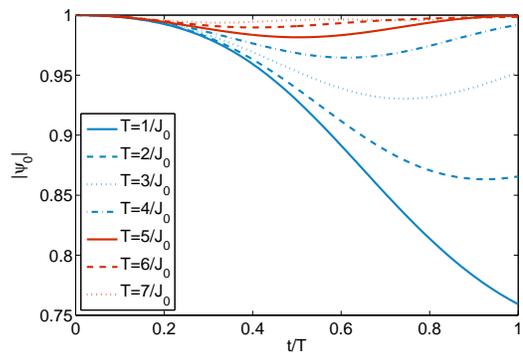}
\caption{ $|\psi_0(t)|$ vs dimensionless time $t/T$ for different passage times $T$. It approaches to the adiabatic limit $|\psi_0(t)|=1$ for larger $T$.} \label{cc}
\end{figure}

Equations (\ref{ME1}) and (\ref{Pt}) are the primary results to be used in analyzing adiabatic dynamics and passages. To this end, let us first consider a model that is widely used in quantum adiabatic algorithms,
\begin{equation}\label{model1}
H(t)=J_0[\frac{t}{T}\si_x+(1-\frac{t}{T})\si_z],
\end{equation}
where $\si_z$ and $\si_x$ are Pauli matrices. The propagator is
\begin{equation}\label{Ptc}
g(t,s)=\frac{4e^{i\int_s^tE(s')ds'}}{T^2E^2(t)sdE^2(s)}.
\end{equation}
If the system is initially in one of the eigenstate of $H(0)=J_0\si_z$, $|\psi(0)\ra=|E_0(0)\ra=|\uparrow\ra$ ($|\downarrow\ra$), it will adiabatically evolve into the eigenstate $|\psi(T)\ra=|E_0(T)\ra=\frac{1}{\sqrt{2}}(|\uparrow\ra+|\downarrow\ra)$ ( $\frac{1}{\sqrt{2}}(|\tiny\uparrow\ra-|\,\downarrow\ra)$) of  $H(T)=J_0\si_x$ at $T\rightarrow\infty$. The recent efforts have been denoted to finding shortcuts to adiabaticity between the initial state and final state (e.g., see \cite{Muga} and references therein). While these shortcuts may be realized by some fast-varying Hamiltonians that are different from the adiabatic ones, our formalism can achieve the desired shortcuts using the same Hamiltonians but with different passage times. Fig.~ (\ref{cc}) depicts $|\psi_0(t)|$ vs the dimensionless time $t/T$ for different passage times $T$, calculated by Eqs.~(\ref{ME1}) and (\ref{Ptc}). It should be noted that a transition from $|\uparrow\ra$ to $\frac{1}{\sqrt{2}}(|\tiny\uparrow\ra+|\,\downarrow\ra)$ does not show up if $1/T>J_0/5$, meaning that there is no shortcut to adiabaticity for a short period $T$. It is interesting to note, however, when $1/T\le J_0/5$, the shortcuts to adiabaticity can be realized, which can be the upper bound for the time changing rate of these shortcuts in this specific model.

{\em Noise induced Adiabaticity.}---A suprising phenomenon can be observed when noise is added to our one-component Eq.~(\ref{ME1}). As a random and fast-varying function, noise is typically a source of destruction that may cause decoherence. We show that a certain type of noise surprisingly induce adiabaticity. Technically, we will show that noise can render the adiabaticity  condition valid in the same way as the fast-varying function in $g(t,s)$. Crucial to our investigation of this issue is to find an appropriate physical model which can incorporate the required white noise. It turns out that the simplest physics model
is a two-state dephasing model, which contains a white noise modifying the strength of a Hamiltonian in a random manner.  Such a model can be easily obtained if we replace $J_0$ with $J_0+c(J,W,t)$ in Eq.~(\ref{model1}).  Here $c(J,W,t)$ is a white noise, more specifically, it is the biased Poissonian white shot noise \cite{Peter1,Peter2}. Note that $J$ is the noise strength and $W$ measures the average frequency of noise shots ( if not specifically mentioned, the `noise' always refers to the white noise throughout this paper). When $W$ goes to infinity, $c(J,W,t)$ becomes to a continuous-time white noise denoted as $c(J,t)$. Note that the noise term only rescales the eigenvalues $E_m$'s to $[1+c(J,t)/J_0]E_m$ but does not change the instantaneous eigenstates. Physically, the noise model considered here naturally arises in many physically interesting settings such as a rotating spin that is subjected to a random magnetic field. 

\begin{figure}[htbp]
\centering
\includegraphics[width=2.8in]{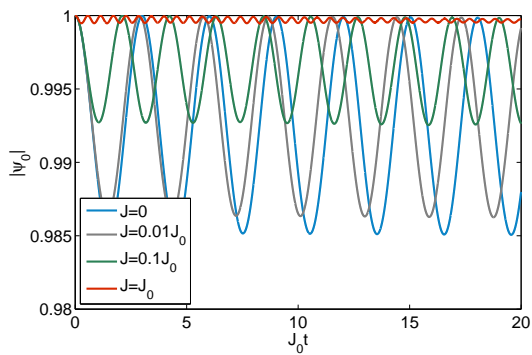}
\caption{Model A: $|\psi_0(t)|$ for different noise strengths. $\Om=0.4J_0$ and $\om=J_0$ in the adiabatic regime. }\label{ca0}
\end{figure}

\begin{figure}[htbp]
\centering
\includegraphics[width=2.8in]{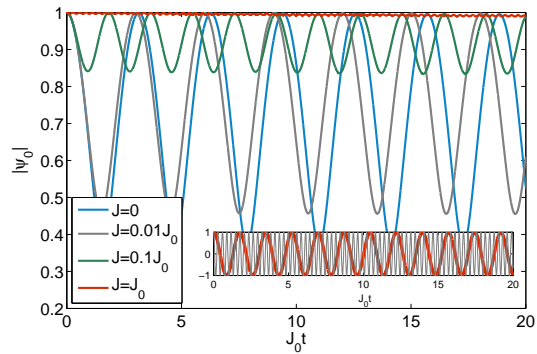}
\caption{ Model A: $|\psi_0(t)|$ for different noise strengths. $\Om=5J_0$ and $\om=5J_0$ far from the adiabatic regime. Inset plots the real part of fast-varying factor $e^{-i\int_0^sc(J,s')kds'}$ in grey curves and the slowly-varying $e^{-i\int_0^s[J_0k+\Om\sin^2\alpha(s')]ds'}\psi_0(s)$ in red curves. }\label{ca}
\end{figure}

{\em Model A: a single TLS model.}---The most general time-dependent Hamiltonian of a TLS or qubit may be written as $H(t)=J_0(a\si^x+b\si^y+\om\si^z/2)$, where $a$ and $b$ describe the transverse fields and $\om$ is the longitudinal fields. The instantaneous eigenstates of $H(t)$ can be expressed by
\begin{eqnarray}\non
|E_0\ra&=&e^{-i\beta}\cos\alpha|\uparrow\ra+\sin\alpha|\downarrow\ra, \\ \label{eS}
|E_1\ra&=&-e^{-i\beta}\sin\alpha|\uparrow\ra+\cos\alpha|\downarrow\ra,
\end{eqnarray}
where $\beta=\tan^{-1}(b/a)$ and $\alpha=\cos^{-1}\frac{k+\om}{\sqrt{2k^2+2k\om}}$ with $k\equiv\pm\sqrt{\om^2+4a^2+4b^2}$.

We now consider a simple case with $a=\cos(\Om t)$, $b=\sin(\Om t)$ where $\Om$ is a constant frequency. The propagator now is
\begin{equation}\label{Pt2}
g(t,s)=\frac{\Om^2}{k^2}e^{i\int_s^t[E(s')+\Om\sin^2\alpha(s')]ds'},
\end{equation}
where $E(s')=[J_0+c(J,s')]k$. The model physically describes a spin-$1/2$ particle driven by a periodical magnetic field. As an example, if $\Omega=\omega$ and in the rotating framework, the total Hamiltonian is $J_0\sigma_x+J\sigma_x$, {\em which is a typical dephasing model and may be accessible experimentally}. When $\Om$ approaches to zero, the standard adiabaticity can be reached, which is shown by the blue curve in Fig. \ref{ca0}. Surprisingly, even when there exists a weak noise, adiabaticity is improved rather than 
destructed as shown by the other curves. In the adiabatic regime, it is shown that the stronger the noise is, the better adiabaticity is achieved. It is in stark contrast to our common understanding on how noise affects adiabaticity where noise is a source of disorder or a nuisance. A more surprising result is that, in the non-adiabatic regime noise can even induce adiabaticity. Strong noise can bring about a system from a non-adiabatic regime into a adiabatic regime.

Consider the non-adiabatic regime where $\Om=5J_0$ and $\om=5J_0$. In Fig. \ref{ca}, the blue curve depicts the noise free term $|\psi_0|$ which strongly oscillates from $1.0$ to a minimum $~0.36$. The system undergoes transitions between $|E_0\ra$ and $|E_1\ra$. the other  curves show that $|\psi_0|$ can be decreased by increasing the noise 
strength $J$. For a weak noise with $J=0.01J_0$ shown in the grey curve, the minimum of $|\psi_0|$ is increased to $~0.46$. When the noise is moderate ($J=0.1J_0$), the green curve shows the minimum becomes $~0.85$. When noise has $J=J_0$, it induces the perfect adiabaticity as shown in the red curve.

It is worth emphasizing that the above new phenomenon on the noise-induced-adiabaticity is a remarkable example showing that noise can play a positive role in inducing adiabaticity in a very simple system that may arises spontaneously in many contexts in physics. It reveals an interesting observation that the adiabatic process can be realized in quantum open systems in a way that is not seen in a closed quantum system.
 %he process from a system in the non-adiabatic regime.

\begin{figure}[htbp]
\centering
\includegraphics[width=2.8in]{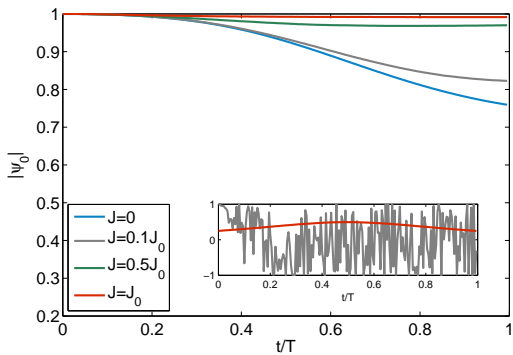}
\caption{Model B: $|\psi_0(t)|$ for different noise strengths.  $T=1/J_0$ far from the adiabatic regime. The noise-free dynamics is also shown in Fig. \ref{cc}.  Inset plots the real part of the fast-varying $e^{-i\int_0^sc(J,s')k(s')ds'}$ in grey curves and $e^{-i\int_0^sJ_0k(s')ds'}\psi_0(s)/k^{2}(s)$ in red curves. }\label{cb}
\end{figure}

{\em Model B: a double TLS system.}---Consider two coupled TLS systems embedded in their individual baths,
\begin{equation}
H = J_0(c\, \si_1^{+}\si_2^-+c^*\si_1^-\si_2^{+}+B_1 \si_1^z+B_2 \si_2^z)
\end{equation}
where $c=a-ib$ and $\om$ are noise-free parameters. $B_{1}=B+\om/4$ and $B_{2}=B-\om/4$, where $B$ is a noise but $\om$ as a difference between $B_1$ and $B_2$ is noise-free. Physically, the two TLSs are subject to a collective noise but to different external fields. When the system state is initially at a single-exciton state: $|\psi(0)\ra=\mu|\uparrow\downarrow\ra+\nu|\downarrow\uparrow\ra$, $|\mu|^2+|\nu|^2=1$, the effective Hamiltonian for this model could be written as $H_{\rm eff}=J_0[(c\,\si_1^+\si_2^-+h.c.)+\om(\si_1^z-\si_2^z)/4]$. The corresponding eigenstates of $H_{\rm eff}$ could be also expressed by Eq.~(\ref{eS}) if $|\uparrow\downarrow\ra$ and $|\downarrow\uparrow\ra$ are mapped into the two states for a TLS,  $|\uparrow\downarrow\ra\Rightarrow|\uparrow\ra$ and $|\downarrow\uparrow\ra\Rightarrow|\downarrow\ra$. The operator mapping is $\si_1^+\si_2^- \Rightarrow \si_{+}$ and $(\si_1^z-\si_2^z)/2 \Rightarrow \si_z$. We now use the same parameters as in the first example, $a=\frac{t}{T}$, $b=0$, and $\frac{\om}{2}=1-\frac{t}{T}$. The propagator is
\begin{equation}\label{Ptc1}
g(t,s)=\frac{4e^{i\int_s^t[J_0+c(J,s')]k(s')ds'}}{T^2k^2(t)k^2(s)},
\end{equation}
where $k(t)=2\sqrt{T^2-2tT+2t^2}/T$. As in the first example our reference is an adiabatic passage from an eigenstate $|\uparrow\downarrow\ra$ of $H(0)=J_0(\si_1^z-\si_2^z)/2$ to  $|\uparrow\downarrow\ra+|\downarrow\uparrow\ra$ of $H(T)=J_0(\si_1^+\si_2^-+h.c.)$. Generally, the basic conclusions will remain intact if we use different types of $H(0)$ and $H(T)$. Now $H_{\rm eff}$ for the two coupled TLSs is equivalent to that in the first example, though representing different physics.

We consider non-adiabatic regime where $J_0=1/T$. In Fig.~\ref{cb}, the blue curve depicts the noise free $|\psi_0|$, which is the same as the corresponding curve in Fig.~\ref{cc}. Again, the other curves show the onset of the adiabaticity induced by noise. The probability for the system to become $|E_0(T)\ra$ is increased with the noise strength. Physically, it means that while noise keeps the system on the eigenstate of $H(T)$, hence it induces adiabaticity from the non-adiabatic regime, with a much shorter period $T$. Fig.~\ref{cb} also shows that creation of adiabaticity does not even require strong noise strength. For instance, for $J=J_0$, $|\psi|$ is already maintained as high as above $0.99$ and $T$ is $8$ times faster than the same passage in the adiabatic regime (see, Fig.~\ref{cc}).

It is interesting to note that this specific system {\em suffers} from two types of noises. The first type of noise, characterized by $B$, acts on a time-dependent decoherence-free subspace (DFS) \cite{Viola,Lidar2}, hence gives rise to no effect on dynamics. The second noise, embedded in the strength of the Hamiltonian, induces adiabaticity. 
This discovery may open up a new way of applications in quantum information science and technologies such as  holonomic and adiabatic quantum computation.

{\em Discussions.}---Adiabaticity is shown to be achievable through introducing an external white noise, which can significantly modify the integral term contained in Eq.~(\ref{ME1}). Specifically, if $g(t,s)$ is a fast-varying time-dependent noise function whereas  $\psi_0(s)$ is slow, the integral will vanish. Our noise model shifts the strength $J_0k(s')$ to $[J_0+c(J,s')]k(s')$ in the oscillation function $e^{-i\int_0^sE(s')ds'}$ in Eqs.~(\ref{Pt}), (\ref{Ptc}) and (\ref{Pt2}). Insets in Fig.~(\ref{ca}) and (\ref{cb}) plot the fast-varying factor and the other factor in the integral and show how noise washes out the accumulation of the slow function.  We also show that the angle in $\psi_0(t)$ goes to zero in both noise-induced adiabatic and adiabatic regimes. {\em It should be emphasized that similar induced adiabaticity occurs if we apply a control field that provides the fast-varying factor in $g(t,s)$}. %The difference is that the adiabatic process induced by control fields time-evolves unitarily.  

When the dynamics approaches to an adiabatic regime $|\psi_0(t)|\approx1$, the quantum sate evolves on the eigenstate of $H(t)$, $|\psi(t)\ra\approx|E_0(t)\ra$. More precisely, the standard stochastic dynamics gives the system density matrix $\rho(t)$ via $\rho(t)=M[|\psi(t)\ra\la \psi(t)|]$. $\rho(t)\approx |\psi_0(t)|^2 |E_0(t)\ra\la E_0(t)| \approx |E_0(t)\ra\la E_0(t)|$ holds only if $|\psi_0(t)|\approx1$. Generally, the dephasing white noise will drive the system to a mixed state where the off-diagonal matrix elements will vanish (whereas adiabaticity induced by control fields time-evolves unitarily).
%\section{Conclusion}\label{con}

{\em Conclusion.}---We employ the Feshbach P-Q partitioning technique to derive a one-component integro-differential equation, which naturally gives rise to adiabatic condition. Moreover, such one-component dynamical equation can be used to demonstrate the onset of  adiabaticity induced by the white noise. 
We work out two examples by analyzing the adiabatic conditions and numerically exhibiting the noise effect on adiabaticity. In addition, we show the significant reduction on the passage time to adiabaticity. 

%The one-component dynamic equation enables us to understand the interesting phenomenon in terms of the propagator. If the propagator has a fast-varying factor, the integral in the equation will sum up as zero. Subsequently, the system will stick on its instantaneous eigenstate. The noise induction mechanism has been examined by two different examples. We argue that the induced adiabaticity may occur in nature.
%In addition, 
The  new discovery can be applied to many ongoing physical implementations of quantum information and quantum computing protocols such as adiabatic quantum computing and the fast energy transfer. As one particular application, we point out that the time evolution required by an adiabatic quantum algorithm can be 
significantly speeded up by an engineered external noise. 
% of our type or even artificially implemented noise.

{\em Acknowledgments.}--- We thank J. G. Muga for his useful comments. We acknowledge grant support from the NSFC Nos. 11175110, 91121015, and  11005099, the Basque Government (grant IT472-10), the Spanish MICINN ( No. FIS2012-36673-C03- 03)  and the NSF PHY-0925174, AFOSR No.~FA9550-12-1-0001.

\end{document}